\documentclass[usenatbib,usegraphicx]{mn2e}
\usepackage{times}
\include{epsf}

\title[Simulating Late Type Galaxies]{Stellar Halo Constraints on Simulated Late Type Galaxies }
\author[Chris B. Brook, Daisuke Kawata, Brad K. Gibson \& Chris Flynn]
       {Chris B. Brook$^1$, Daisuke Kawata$^1$ Brad K. Gibson$^1$\& Chris Flynn$^2$\\
        $^1$Centre for Astrophysics \& Supercomputing, 
        Swinburne University, Mail \#31, P.O. Box 218,
	Hawthorn, Victoria, 3122, Australia\\
$^2$Tuorla Observatory, Piikki\"{o},
FIN-21500, Finland}

\date{Accepted MNRAS}

\pagerange{\pageref{}--\pageref{}}
\pubyear{2003}

\begin{document}

\maketitle

\label{firstpage}

\begin{abstract}
How do late type spiral galaxies form within the context of a  CDM cosmology?  We contrast N-body, smoothed particle hydrodynamical simulations of galaxy formation which employ two different supernova feedback mechanisms. Observed mass and metallicity distributions of the stellar halos of the Milky Way and M31 provide constraints on these models. A strong feedback model, incorporating an adiabatic phase in star burst regions, better reproduces these observational properties than our comparative thermal feedback model. This demonstrates the importance of energy feedback in regulating star formation in small systems, which collapse at early epochs, in the evolution of late type disk galaxies.   

\end{abstract}
\begin{keywords}
galaxies: formation --- evolution --- halo --- disk
\end{keywords}

\section{Introduction}
Formation scenarios for the Milky Way  have been dominated by  two canonical models: the rapid collapse of a proto-galactic cloud (Eggen, Lynden-Bell \& Sandage 1962), and the hierarchical build up from low mass fragments (Searle \& Zinn 1978). Burkert (2001) reviews evidence that the early dissociation of gas from dark matter in the first stages of the hierarchical build up allows these two models to be  combined into one consistent formation scenario. In this scenario, the Galactic spheroid is formed by hierarchical merging of substructures, while the disk components result from the smooth infall of diffuse gas. This picture is consistent with the predictions of classical  cold dark matter (CDM) scenarios (White \& Frenk 1991; Kauffmann et~al. 1993) 

Simulations now show that the observed large scale properties of galaxies
   can be reproduced within cold dark matter (CDM) cosmological models. For example, Steinmetz \& Muller (1995) studied the chemo-dynamical evolution of disk galaxies, and succeeded in distinguishing the chemical properties between halo, disk and bulge stars. Raiteri, Villata, \& Navarro (1996) and  Berczik (1999), reproduced the correlation between [O/Fe] and [Fe/H] of solar neighbourhood stars  by tracing the metal enrichment by Supernovae Type~II (SNe~II) and Type~1a (SNe~Ia). Steinmetz \& Navarro (1999), Navarro \& Steinmetz (2000), and Koda, Sofue, \& Wada (2000) discussed the zero point of the Tully-Fisher relationship by analysing the end products of their chemical evolution and stellar population synthesis models. The observed eccentricity-metallicity relation of Milky Way halo stars  has also been successfully reproduced (Bekki \& Chiba 2000; Brook et~al. 2003).

However, several problems continue to plague these simulations.  Famously, the resulting structures are deficient in angular momentum (Navarro, Frenk \& White 1995). These simulations also fail to create galaxies in which most baryonic matter resides in the thin disk, as observed in the Milky Way. Typically, these simulations result in a stellar halo which is as massive or more massive than the stellar disk. For example, in a recent high resolution study, Abadi et~al. (2003, ANSE-I hereafter) create a galaxy with a stellar halo which contains over 60\% of the total stellar mass of the system, and a thin disk which constitutes only 17\% of the stellar mass, more akin to an S0 than a late type disk galaxy.  Such  massive stellar halos inevitably  have a greater metallicity  than observed in late type spirals.    This leaves open the question of the formation processes which lead to late type disk galaxies such as the Milky Way, within the context of  hierarchical structure formation scenarios.  

It has long been postulated that energy feedback from SNe is necessary to solve many of the above problems.
Unfortunately, energy feedback from SNe explosions remains one of the most difficult, yet most critical, processes to model in galaxy formation simulations, largely because feedback occurs on sub-resolution scales. A number of different SNe feedback implementations have been employed in chemo-dynamical codes;  for a survey of such methods, see Thacker \& Couchman (2000, TC hereafter). 

Katz (1992) smooths thermal energy into nearest neigbour gas particles using the smoothed particle hydrodynamic (SPH) kernel. This feedback scheme  is known to be largely inefficient, as the characteristic timescale of radiative cooling in the high density regions, where star formation typically occurs, is shorter than the dynamical timescale.
 Navarro \& White (1993) extend this technique to include both thermal and kinetic energy. A parameter, {\it{f}}, controls the fraction of energy input as kinetic as opposed to thermal energy. Although {\it f} and the total amount of energy both may vary, this remains a commonly employed feedback scheme in galaxy formation simulations (e.g.  Steinmetz \& Navarro 2002; Kobayashi 2002; Nakasato \& Nomoto 2003; ANSE-I). 
  Representations of the multi-phase interstellar medium are made by Hultman and Pharasyn (1999), and by Springel \& Hernquist (2003), but these studies are limited by the two phases not being  dynamically independent. In Gerritsen (1997), energy is returned by heating an individual SPH particle, a technique which is successful in dwarf size systems. TC and Kay et~al. (2002) examine a scheme in which the feedback region is made adiabatic. Such a feedback scheme was shown to be more effective in an isolated, dwarf type galaxy than in a Milky Way sized galaxy. 

We examine in this paper the link between the energy feedback and the properties of the final galaxy; specifically their halo/disk stellar mass ratio and halo metallicity distribution function (MDF). We demonstrate the importance of  feedback in regulating star formation in  the small systems that collapse at early epochs of galaxy evolution.   
In section~2, details of our chemo-dynamical code, {\tt{GCD+}}, and our two models incorporating different feedback schemes are provided.   We then examine and compare the properties of the two simulated galaxies in section~3, concentrating on the halo/disk stellar mass ratio and the halo MDFs of the simulated disk galaxies. One  feedback model results in a galaxy which is a better representation of a late type galaxy than the other. In section~4 we trace the baryons which form the disk of the galaxies, highlighting the processes which lead to the difference in the fraction of stars ending up in the disk compared to the halo. The discussion in section~5 centres on the insights gained into the manner in which thin disks form from structures within CDM cosmologies.      

\section{The Code and Model}
Our Galactic chemo-dynamical code, {\tt{GCD+}}, self-consistently models the effects of gravity, gas dynamics, radiative cooling, and star formation. We include SNe~II and SNe~Ia feedback, and avoid use of the instantaneous recycling approximation in modeling the chemical enrichment. Details of {\tt{GCD+}} can be found in Kawata \& Gibson
(2003). We describe here only the details concerning the inclusion of SNe energy feedback. 

We assume that 10$^{51}$ergs is fed back as thermal energy from each SNe. In our {\it {thermal feedback model}}, denoted ``TFM'' hereafter, energy from SNe~II and SNe~Ia is smoothed over surrounding gas particles according to the SPH kernel, in the form of thermal energy, as in Katz (1992). 
 
Our comparative model incorporates a different feedback mechanism for SNe~II. These SNe trace starburst regions, and hence suffer the most from the inefficiency of the feedback algorithms used in the TFM. In our {\it {adiabatic feedback model}}, ``AFM'' hereafter, gas within the SPH smoothing kernel of SNe~II explosions is prevented from cooling. This adiabatic phase is assumed to last for the lifetime of the lowest mass star which ends as a SNe~II, i.e. the lifetime of an $8$ M$_\odot$ star ($\sim$100 Myr). In the AFM, the energy released by SNe~Ia, which do not trace starburst regions, is fed back as thermal energy. The AFM is similar to a model presented in TC. 

We base our semi-cosmological simulations  on the galaxy formation model of Katz \& Gunn (1991).  The initial condition is an isolated sphere of dark matter and gas, onto which small scale density fluctuations are superimposed, parameterized by $\sigma_8$. These are set up using grafic in COSMICS, which is the predecessor of grafic2 (Bertschinger 2001). These perturbations are the seeds for local collapse and subsequent star formation. Solid-body rotation corresponding to a spin parameter, $\lambda$$\equiv$$J|E|^{1/2}/ GM_{tot}^{5/2}$, is imparted to the initial sphere, where $J$, $E$ and $M_{tot}$ are the total angular momentum, energy, and mass of the system, respectively. A low value of
$\lambda$ genererally results in an elliptical 
galaxy, while a large value of $\lambda$, in the 
absence of late major mergers, tends to result in a disk like galaxy (Kobayashi  2002). 
We chose spin parameter, $\lambda$=0.0675, and initial conditions 
in which no major merger occurs at late epochs, because we are interested
in the formation of Milky Way type disk galaxies. Other relevant parameters include the total mass (5$\times$10$^{11}$M$_\odot$), baryon fraction ($\Omega_{b}$=0.1), star formation efficiency $c_*$=0.05, $\sigma_8$=0.5. We employed 38911 dark matter and 38911 gas/star particles, making the resolution of this study comparable to other recent studies of disk galaxy formation (e.g. ANSE-I).

We again emphasise that the two models analysed here are identical, except in the way in which the energy from SNe~II affects the surrounding gas.

\begin{figure*}
\epsfxsize=13.0cm \epsfbox{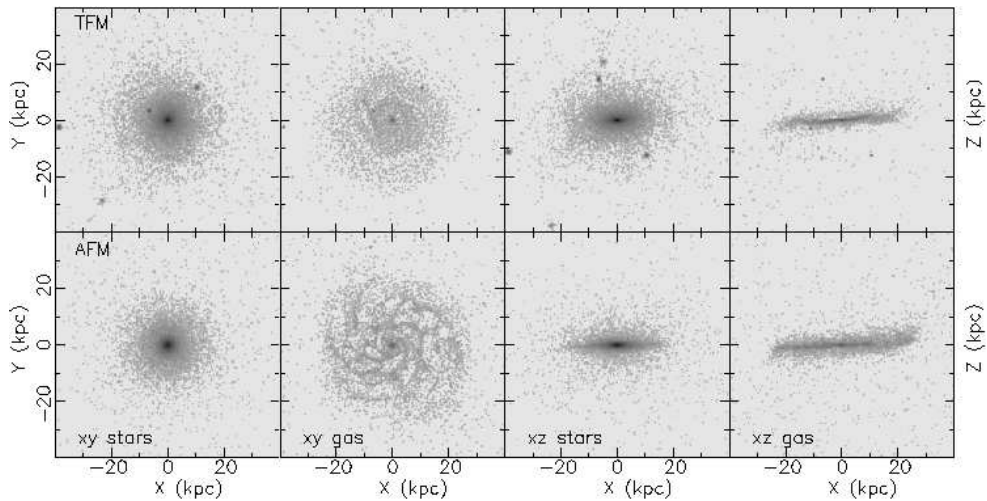}
\caption{Density plots for TFM (top) and AFM  at $z$=0. Stars and gas are shown  face-on (XY plane) and edge-on (XZ). In the TFM, a thin disk is evident in the gas distribution, but the galaxy's stellar mass is dominated by halo stars, and resembles an S0 galaxy. In the AFM,  the galaxy is dominated by a thin stellar disk. A large gaseous thin disk, still undergoing star formation, has also formed.   \label{fig1}}
\end{figure*}

\section{Results}

Fig.~1 shows a density plot of the final stellar and gaseous  components  of the two models. Both simulations result in flattened stellar structures, and thin gaseous disks.  
In the TFM, the stellar mass is dominated by halo stars, and a thin disk is evident in the gas distribution. The final galaxy more closely resembles an S0 galaxy than a late type spiral. Star formation is rapid during early epochs (Fig.~2), as a result of the inefficiency of the feedback. The peaks in the star formation rates are related to early major merger episodes. Star formation continues at a steady rate, primarily in the disk, over the last $\sim$7 Grys, a relatively quiescent period with only a few minor merger events.  

It is clear from the edge on view of the stellar populations in Fig.~1 that the AFM has created a more dominant stellar disk.
 Star formation in the AFM is suppressed at early epochs (Fig.~2) when compared to the TFM. The signatures of the early mergers of the TFM are also seen in the AFM star formation peaks, a result of the identical initial conditions.  At later times the greater availability of gas means the AFM creates more stars, $\sim$1.9$\times$10$^{10}$M$_\odot$, than the TFM, $\sim$8.2$\times$10$^{9}$M$_\odot$, in the last 8 Gyrs. The  result is the creation of a more dominant  stellar disk. A large gaseous thin disk, still undergoing star formation, has also formed. The mass of cold gas (T$<$5$\times$10$^4$K) associated with the disk is 
larger for the AFM, $\sim$8.2$\times$10$^{9}$M$_\odot$  than the TFM,  
$\sim$4.5$\times$10$^{9}$M$_\odot$. 
 
In this study, we focus on the mass and metallicities of the stellar halos of our simulated galaxies. In defining the components of the galaxy, the important issues for the purposes of this study are 
1) that the halo and disk stars are defined the same way in each model, allowing a comparison, and
2) that these  definitions do not influence the main result highlighted. 
We define halo stars  spatially using two hemispheres defined by 4$<$$R$$<$50 kpc and  $|Z|$$>$4 kpc. The density gradient in shells of these hemispheres is integrated between 4 and 50 kpc to find the mass of the stellar halo. Disk stars are defined by using the  distribution of specific angular momentum ($Lz_s$) as a function of specific binding energy($E_s$). It is clear from Fig.~1 that the gas particles are predominantly in a disk. These disk gas paricles are  confined to a small region in  $Lz_s$ vs. $E_s$ space. This same region is used to define simulation disk stars; cuts in the $Z$ components of velocity ($|V_Z|$$<$65 km/s) and position ($|Z|$$<$1 kpc) have also been imposed.

\begin{figure}
\epsfbox{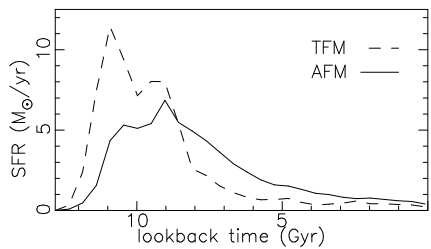}
\epsfxsize=5.6cm
\caption{Global star formation rate (SFR)
as a function of lookback time for the two supernova energy feedback
models applied to the same initial conditions. Inefficient feedback of the TFM results in high SFR at early times compared to the AFM.}
\label{f2}
\end{figure}

The stellar population of the TFM is dominated by halo stars; the stellar halo  being a factor of 1.2 times more massive than its stellar disk population. The AFM has a  halo/disk mass fraction $\sim$6 times smaller, more closely resembling a late type galaxy. This is still close to an order of magnitude larger than the halo/disk ratio of the Milky Way. The large mass of the stellar halo in the TFM is intimately linked to its high metallicity, as seen in Fig.~3.
The peak of the halo MDF has shifted from [Fe/H]$\sim$0 for the TFM (dashed line),  to [Fe/H]$\sim$$-$1.0 for the AFM (solid line), a full factor of ten in metallicity.  The peak of the halo MDF in the AFM is somewhat higher  than that of the Milky Way halo MDF, which peaks at [Fe/H]$\sim$$-$1.5 (Ryan \& Norris 1991), but is  lower than the peak for M31, [Fe/H]$\sim$$-$0.7 (Durrell et~al. 1994). These results suggest that strong feedback from SNe is important for the formation of the stellar halos of disk galaxies.
The results were found to be  robust when we  employed different definitions of halo and disk stars, incorporating spatial information, kinematic information, age information or combinations of these.   

Significant progress has been made recently on the theoretical and observational signatures of satellite accretion in present day stellar populations, e.g. Helmi \& White 2000; Bekki \& Chiba 2001; Brook et~al. 2003. Further, the  model used in  TC which incorporated adiabatic feedback was shown to have a greater effect on smaller systems. Thus, the natural place to search for the formation processes leading to the different galaxy properties produced by the different models was in the subclumps of the simulation.

\begin{figure}
\begin{center}
\epsfbox{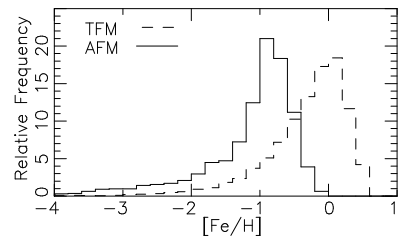}
\epsfxsize=5.16cm 
\caption {Halo metallicity distributions for the two models, giving the relative fraction of halo stars within a given metallicity bin. Halo stars are defined  by weighting star particles by mass within the regions 4$<$$R$$<$50 kpc and  $|Z|$$>$4 kpc. 
\label{fig 3} }
\end{center}
\end{figure}

\begin{figure*}
\epsfxsize=13.0cm
\epsfbox{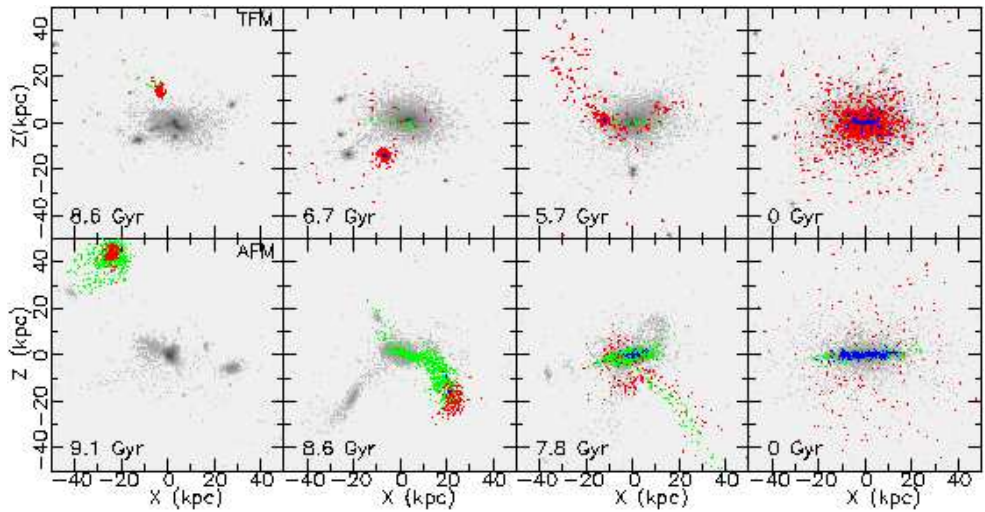}
\caption{The disruption process for a typical satellite from the two models. Red particles are stars present in the satellite prior to any disruption. Gas particles are green. Star particles formed after the disruption process begins are blue. Four edge-on snapshots of the model are shown for these typical merging events.
In the TFM (upper panels), gas falls to the centre of sub-clumps, rapidly forming stars. The stars are accreted onto the halo of the galaxy. Gas from the satellite falls to the disk region of the host galaxy, where new stars are born.  
In  the AFM,  gas remains less densely concentrated than stars and is preferentially stripped, becoming decoupled from the dark matter. The stripped gas accretes smoothly to the disk region. Stars from the satellite are accreted into the halo of the galaxy by $z$=0. Pre-enriched gas stripped from the satellites feeds the thin disk, where new stars (blue) are born over the past $\sim$8 Gyrs.} 
\end{figure*}

In Fig.~4, we show the disruption process for a typical  satellite from the TFM and AFM. Red particles are stars present in the satellite prior to any disruption. Gas particles are green. Star particles formed after the disruption process begins are blue. A grey scale density plot of the simulation stars is in the background. There are $\sim$1000 baryon particles in each satellite, corresponding to $\sim$10$^9$ M$_\odot$.
Four timesteps  are shown edge-on, for these typical merging events.

The disruption of a satellite in the TFM is shown in the upper panels of Fig.~4. The satellite begins as a dense, stellar dominated system. This is a result of the rapid early star formtion, between 11 and 8 Gyrs ago (Fig.~5, dashed line). By 5.7 Gyrs ago, accretion is well underway, and star formation has ceased within the dwarf. Due to their collisionless nature, the accreted stars spread throughout the halo by z=0. Gas from the satellite falls to the disk region  of the host galaxy, where new stars (blue) are born. The amount of gas falling into the disk region, and forming stars, is small as most gas has been used up before the satellite merged.

In the AFM, star formation in the subclumps is regulated (Fig.~5, solid line) prior to disruption. Gas remains less densely concentrated than stars and is preferentially stripped. This pre-enriched gas becomes dissociated from the dark matter, and accretes smoothly on to the disk region. On the other hand, stars from the satellite are accreted into the halo of the galaxy by z=0. Therefore, the stars which end up in the halo are the oldest stars formed in the subclumps, as well as the most metal poor, as chemical enrichment has been suppressed in these small systems by the strong feedback in the AFM. 

The accretion events shown in this section are events which occur at intermediate age in the galaxy formation process. They are not meant to imply that late accreting satellites are the source of all disk and halo material. We show them to demonstrate the effectiveness of SNe feedback  on satellites. AFM also leads to regulated star formation in the smaller systems which have collapsed at high redshift, as is evidenced by Fig.~2. In a CDM cosmology, the stellar halo is likely to be built up by the accretion of a significant number of such small systems, including tidally stripped stars during multiple mergers of building blocks at high-redshift (Bekki \& Chiba 2001). Our results demonstrate that the AFM produces enough SNe energy to suppress star formation and chemical enrichment in the building blocks of galaxy formation, and that this leads to a less massive, more metal poor, stellar halo, in better agreement with observation.  

\begin{figure}
\epsfbox{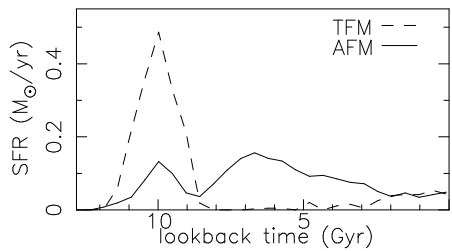}
\epsfxsize=5.588cm
\caption{Total star formation rates (SFR) within the  satellite systems of Fig.~4. As in the host galaxy, star formation in the satellites is early and rapid in the TFM compared to the AFM.
\label{fig5} }

\end{figure}

\section{Discussion}
Energy feedback has emerged as a possible solution to several of the problems that plague models of the formation of late type disk galaxies in  CDM scenarios. Despite this recognition, the manner in which feedback is incorporated into SPH  codes remains problematic.  In this study we have been concerned primarily with the effects of feedback on the mass and metallicity distribution of the stellar halos of simulated galaxies, and the effects it has in regulating star formation in high-redshift galactic building blocks.

We employ two different feedback schemes into a seed galaxy with the same initial perturbation.
Incorporating thermal feedback similar to that of Katz (1992) results in a flattened, disk like galaxy which  resembles those of previous studies (e.g. Steinmetz \& Mullar 1995; Berzcik 1999; ANSE-I). The fundamental problem
  is that, compared to the Milky Way, the ratio of the mass of the halo
  and disk components is much too high, by about two orders of magnitude, and the metallicity of the halo
  component is about one order of  magnitude too high. The
 mismatch with M31 observations is less, but still significant.  In the TFM, the gas cools to the centre of building blocks, rapidly forming stars (Fig.~2). The rapid star formation cycles also result in metallicities rapidly approaching $\sim$solar levels and more stars with high metallicity. Due to the collisionless nature of the stellar component, these stars end up in the halo of the host galaxy after tidal disruption. The halo thus becomes very massive, and very metal rich.

With the AFM, gas is slowed from cooling in star burst regions. 
Star formation is regulated in small systems, such as building blocks, and lower fractions of the initial baryonic content of these building blocks are converted to stars. These building blocks thus have lower metallicity stars and a hotter, more diffuse gas content, allowing gas to be dissociated from the dark matter, whose gravitational collapse ultimately drives structure formation. During the tidal disruption of these building blocks, the gas is preferentially stripped and, due to its dissipative nature, accretes smoothly onto the thin disk; this pre-enrichment helps in alleviating the G-dwarf problem (Fenner \& Gibson 2003). The stars which formed in these building blocks are accreted preferentially into the halo of the galaxy. As a smaller fraction of baryons are turned into star particles in the building blocks in  the AFM, the halo of the resulting host galaxy is less massive and more metal poor. Also, even if they are accreted into the galaxy at a later epoch than the gas, the stars stripped from these building blocks are older than the stars which subsequently form in situ in the thin disk: halo stars are old and metal poor, disk stars are new and metal rich. The result is a disk galaxy which is a better realisation of the Milky Way and M31, compared to the canonical TFM.    

We note here that several other techniques  were employed in an attempt to reproduce a halo mass and MDF closer to those observed in the Milky Way. These included varying the star formation efficiency, adjusting the amount of feedback incorporated as kinetic energy, and incorporating ``re-ionisation epochs'' during which radiative cooling was essentially ``squelched''(Somerville 2002). None of these adjustments were successful in lowering the mass and metallicity of the simulated stellar halo. To date, only the AFM has succeeded in significantly lowering the metallicity and mass of the simulated galaxies stellar halo. 
The success of the AFM highlights two features which a feedback scheme must pocess in sufficient measures: an ability to regulate star formation at early epochs (Fig.~2), and an effect which is more pronounced in smaller systems (the early collapsing subclumps and later forming dwarf galaxies) than in a Milky Way sized system.

Using a feedback scheme similar to the AFM, Thacker \& Couchman (2001), found an improvement in the ``angular momentum problem'' of their simulated galaxy, when compared with feedback schemes which incorporate thermal and kinetic energy. However, their simulation was halted at redshift $z$=1, and thus the important epoch of disk formation was not simulated. ANSE-I argues that the angular momentum problem may be ``evidence that the difficulty in reconciling the properties of the simulated galaxy with those of observed spirals lies in the presence of the dense, slowly rotating spheroid that dominates the luminous stellar component.'' We agree, and believe that the observed properties of the stellar halo/disk mass ratio and halo MDF provide strong, important constraints on disk galaxy formation models. We suggest that many previous studies have inevitably failed to reproduce these properties (even if they were not specifically addressed in these studies). 

Besides observations of the Milky Way's stellar halo mass and MDF, which are the main thrust of this paper, additional Local Group observations provide a measure of support for the processes outlined here. The Magellanic Stream is consistent with the suggestion that gas is preferentially stripped from accreting satellite galaxies (Putman et~al. 1998). The Sagittarius Dwarf, and Sagittarius stream, are totally stellar (Koribalski et~al. 1994; Burton \& Lockman 1999), suggesting that this accreted satellite lost its gas first. The fact that Local Group dwarf spheroidals are gas poor, and preferentially closer to the centre of the Milky Way than gas rich  dwarf irregulars provides further evidence, albeit circumstantial, that gas from satellites is stripped prior to stellar accretion (van den Bergh 1999).  
   
Our simulations emphasise that it is critical for disk galaxy formation in a CDM context that building blocks and later accreting satellites be  gas dominant.
We regard our simulations as a successful realistion of  the scenario sketched by Burkert (2001), which emphasised the need for an early phase of gas heating and dissociation from dark matter in forming the Milky Way within  hierarchical cosmologies.

\section{acknowledgments} 
We appreciate helpful discussions with Mike Beasley and Stuart Gill, and
 thank the referee, Vincent Eke, for comments which have improved the paper.     This study made use of the facilities of the
Victorian and Australian Partnerships for Advanced Computing, and is supported financially
by the Australian Research Council. CBB acknowledges the support of DEST through  an Australian Postgraduate Award. CF thanks the Academy of Finland for financial support through its ANTARES program for space-based research.

\end{document}